\documentclass[aps,showpacs,floats,twocolumn,superscriptaddress]{revtex4-1}
\pdfoutput=1
\usepackage{mathrsfs}
\usepackage{bm,amsbsy,amssymb,amsmath}
\usepackage{caption}
\usepackage{graphics,graphicx,dcolumn,fleqn,epic,eepic,float,tabularx}
\usepackage{multirow,rotate,rotating,color}
\usepackage[utf8]{inputenc}
\newcommand{\figref}[1]{Fig.~\ref{fig:#1}}
\newcommand{\eqnref}[1]{Eq.~(\ref{eq:#1})} 
\definecolor{tuered}{RGB}{214,0,74}
\definecolor{tueblue}{RGB}{0,102,204}

\newcommand{\revisedtext}[1]{{#1}}

\newcommand{\lvci}{\mathbf{c}_i}

\usepackage{tikz,pgfplots}
\usetikzlibrary{external,spy,calc}
\tikzset{external/system call={pdflatex \tikzexternalcheckshellescape -interaction=batchmode --enable-write18 --shell-escape -tok_size=200000000  -jobname "\image" "\texsource"}}
\usepackage{subfigure}
\pgfplotsset{compat=newest}
\usepackage{contour}
\usepackage{lipsum}
\let\ACMmaketitle=\maketitle
\renewcommand{\maketitle}{\begingroup\let\footnote=\thanks \ACMmaketitle\endgroup}
\usepackage{microtype}
\begin{document}
\title{Diffusion dominated evaporation in multicomponent lattice Boltzmann simulations}
\author{Dennis Hessling}
\email{d.m.hessling@tue.nl}
\affiliation{Materials innovations institute (M2i), Elektronicaweg 25, 2628 XG Delft,Netherlands}
\affiliation{Department of Applied Physics, Eindhoven University of Technology, P.O. Box 513, NL-5600MB Eindhoven, The Netherlands}
\author{Qingguang Xie}
\email{q.xie1@tue.nl}
\affiliation{Department of Applied Physics, Eindhoven University of Technology, P.O. Box 513, NL-5600MB Eindhoven, The Netherlands}
\author{Jens Harting}
\email{j.harting@fz-juelich.de}
\affiliation{Helmholtz Institute Erlangen-N\"urnberg for Renewable Energy (IEK-11), Forschungszentrum J\"ulich, F\"urther Stra{\ss}e 248, 90429 N\"urnberg, Germany}
\affiliation{Department of Applied Physics, Eindhoven University of Technology, P.O. Box 513, NL-5600MB Eindhoven, The Netherlands}


\begin{abstract}
We present a diffusion dominated evaporation model using the popular
pseudopotential multicomponent lattice Boltzmann method introduced by Shan and
Chen. With an analytical computation of the diffusion coefficients, we
demonstrate that Fick's law is obeyed. We then validate the applicability of
our model by demonstrating the agreement of the time evolution of the interface
position of an evaporating planar film to the analytical prediction.
Furthermore, we study the evaporation of a freely floating droplet and confirm
that the effect of Laplace pressure is significant for predicting the time
evolution of small droplet radii.
\end{abstract}
\maketitle

\section{Introduction}\label{sec:intro}

Evaporating fluids are ubiquitous in our daily life and in industrial
processes, such as ink jet printing~\cite{Lei2015}, coating~\cite{Nadir2015}
and particle deposition~\cite{Wei2012}.  In particular for suspensions or
polymer solutions, as well as fluids in confined geometries, the evaporation of
individual components can induce fluid flows or a change of relative
concentrations leading to changing rheological and transport properties of the
constituents. For example, the evaporation of a sessile colloidal droplet on a
substrate leads to a capillary flow transporting the colloidal particles to the
edge of a droplet, which finally results in a ring-like
deposit~\cite{Deegan1997}. The ring-like stains can be a useful tool to deposit
particles and can also be disadvantageous when a uniform pattern is desirable.
Another example is the evaporation of droplets on rough or chemically patterned
substrates. Surrounding geometries and the wettability of a substrate have a
large influence on the lifetime of evaporating droplets~\cite{GelMarNai11}.  A
thorough understanding of this impact of evaporation on the fluid behavior is
mandatory to consequently optimize industrial applications and to improve our
fundamental understanding of effects like film formation, droplet drying, or
droplet spreading. 

There are numerous theoretical~\cite{EpsPle50,Popov2005,Stephen2015} and
experimental~\cite{Deegan1997,Marin2011} studies of fluid evaporation.  While
most theoretical studies are limited to the macroscopic scale, experiments
suffer from difficulties that arise by tuning the individual microscale
properties of fluids. The thorough understanding of fluid evaporation calls for
mesoscopic or microscopic details and the flexibility to tune the properties of
individual fluid constituents independently. This is possible by means of
computer simulations. Computer simulations allow access to parameters which are
not easily controllable in experiments and to tune the properties of individual
fluid constituents independently. They can thus help to improve our
understanding of evaporation driven fluid transport. 
Simulations of evaporating fluids often utilize
molecular dynamics (MD)~\cite{Long1996,WangMD2013,Chen2013,Detlef2015}. While
MD offers a very high flexibility in the microscopic details, its computational
cost is very high. Therefore, MD simulations are limited to very small length
and time scales on the nanometre or nanosecond scale~\cite{Detlef2015}. In
order to reach experimentally relevant scales, a continuum approach is more
productive and our method of choice is the lattice Boltzmann method
(LBM)~\cite{Succi2001,BenSucSau92,Raa04}. The LBM has gained popularity for the
simulation of fluid flows due to its straightforward implementation and
parallelization. Soon after its invention, the LBM was extended to simulate
multiple interacting fluid phases and components and today \revisedtext{a plethora of} multiphase
and multicomponent methods exists\revisedtext{~\cite{Liu2016,Shan1993,Gunstensen1991,Swift1995,He1999,Falcucci2011}}. 

\revisedtext{The mesoscale nature of the method combined with the possibility
to add additional fields, external forces, suspended objects, thermal noise, or complex
boundary conditions in a very straightforward manner has made the LBM
particularly popular for applications in microfluidics and soft matter physics.
Many of the physical systems studied in these fields include volatile liquids, where the effect of evaporation plays a dominant role. Therefore, it is not surprising that 
}a number of groups has simulated evaporating fluids using the LBM recently.
Ledesma-Aguilar et al.~\cite{LedVelYeo14,AguVelYeo16} present a diffusion based
evaporation method based on the free energy multiphase lattice Boltzmann method
and demonstrate quantitative agreement with several benchmark cases as well as
qualitative agreement with the experimental data of evaporating droplet arrays.
Jansen et al.~\cite{Jansen2013} study the evaporation of droplets on a
chemically patterned substrate and qualitatively compare the simulation results
with experimental data. Their method is based on a continuous removal of mass
from the droplet and thus does not allow studying transport processes in the
vapor phase.  Yan et al.~\cite{Qing2016} present a thermal model to study the
contact line dynamics during droplet evaporation where the liquid-vapor phase
change is driven by a temperature field and a well defined equation of state.
Joshi and Sun~\cite{Joshi2010} present simulations of drying colloidal
suspensions by means of a modified pseudopotential multiphase model following
Shan and Chen. They assign a fixed mass flux to the system boundary which
causes a reduction in vapor concentration and thus triggers a liquid-vapor
phase change at the interface. However, their results are purely qualitative
and a thorough analytical understanding of the diffusion in the system is
missing.  

In this paper we overcome this limitation and introduce an alternative
evaporation model for the pseudopotential method of Shan and Chen. We focus on
the two-component version of the method~\cite{Shan1993}, but the application to
an arbitrary number of components and the multiphase pseudopotential method is
straightforward. Generally, the pseudopotential LBM is very popular due to its
ease of implementation and flexibility when combined for example with complex
geometries~\cite{Martys1996a,Liu2016}, locally varying contact angles~\cite{HarKunHer06},
or suspended particles~\cite{JanHar11}.  
To trigger evaporation, we do not
impose a mass flux, but instead fix the density of one component at selected
boundary sites which induces a density gradient.  The evaporation process is
diffusion dominated and can be well described using Fick's law with well
defined diffusivities.  We validate the applicability of our model by comparing
the time dependent simulation results of an evaporating planar film and a
freely floating evaporating droplet with their respective analytical
predictions.

This remainder of this paper is organised as follows. Sec.~\ref{sec:method}
introduces the lattice Boltzmann method and our extension for evaporating
fluids. Our results are shown in Sec.~\ref{sec:results} and
Sec.~\ref{sec:final} concludes the article.

\section{Simulation method}\label{sec:method}
\subsection{The lattice Boltzmann method}
%
The lattice Boltzmann equation can be obtained from spatially and temporally discretizing the Boltzmann equation.
Multiple fluid components $c$ are modeled by following the evolution of the single particle distribution function
\begin{eqnarray}
    \label{eq:LBG}
    f_i^c(\mathbf{x} + \mathbf{e}_i \Delta t , t + \Delta t)-f_i^c(\mathbf{x},t)=& - \frac{\Delta t} {\tau^c} [  f_i^c(\mathbf{x},t) \nonumber \\ 
                                                             - f_i^\mathrm{eq}(\rho^c(\mathbf{x},t),\mathbf{u}^c_{eq}(\mathbf{x},t))]
    \mbox{.}
\label{eq:LBM}
\end{eqnarray}
The single particle distribution functions $f_i^c(\mathbf{x},t)$ at positions $\mathbf{x}$ alternatively
stream, as described by the LHS of \eqref{eq:LBM}, along the $i=1,\ldots,19$ discretized directions $\mathbf{e}_i$
and collide, as described by the RHS of \eqref{eq:LBM} at every timestep~$t$.
Throughout this work we utilize $2$ components $c$ and $\overline{c}$.
The collision is achieved by relaxing the probability distribution
functions towards a discretized second-order equilibrium distribution function
\begin{eqnarray} \label{eq:eqdis}
    f_i^{\mathrm{eq}}(\rho^c,\mathbf{u}^c_{eq}) = \omega_i \rho^c \bigg[ 1 + \frac{\lvci \cdot \mathbf{u}^c_{eq}}{c_s^2} - \frac{ \left( \mathbf{u}^c_{eq} \cdot \mathbf{u}^c_{eq} \right) }{2 c_s^2} \nonumber \\
    + \frac{ \left( \lvci \cdot \mathbf{u}^c_{eq} \right)^2}{2 c_s^4}  \bigg]
    \mbox{,}
\end{eqnarray}
where $c_s=\frac{1}{\sqrt{3}}\frac{\Delta x}{\Delta t}$ is the speed of sound and
$\omega_i$ is a weight factor defined as $\omega_{0}=\frac{1}{3}$, $\omega_{1,\ldots,6}=\frac{1}{18}$ and $\omega_{7,\ldots,18}=\frac{1}{36}$.
The densities are defined as  
$ \rho^c(\mathbf{x},t) = \rho_0 \sum_if^c_i(\mathbf{x},t)$, where $\rho_0$ is a reference density, and the velocities are defined as $\mathbf{u}^c(\mathbf{x},t) = \sum_i  f^c_i(\mathbf{x},t) \mathbf{c}_i/\rho^c(\mathbf{x},t)$, 
while the velocity in the equilibrium distribution function is $\mathbf{u}^c_{eq} = \sum_c \rho^c u^c/ \sum_c \rho^c$. 

For brevity and numerical efficiency we choose the lattice constant $\Delta x$,
the timestep $ \Delta t$, the unit mass $\rho_0$ and the relaxation time
$\tau^c$ to be unity, which leads to a kinematic viscosity $\nu^c$ $=
\frac{2\tau-1}{6}=\frac{1}{6}$ in lattice units.

The system boundaries are treated as periodic boundaries by 
    default.  To do so, fluid leaving one system boundary reenters the opposite 
    side and forces are computed across these periodic boundaries. To inhibit 
    flow walls can be constructed by inverting velocities at selected boundary 
    sites~\cite{Succi2001}.

\subsection{The pseudopotential multicomponent lattice Boltzmann method}
For the fluid components introduced above to become immiscible, Shan and Chen
introduced a pseudopotential interaction force
\begin{equation}
    \label{eq:sc}
    \mathbf{F}^c(\mathbf{x},t) = -\Psi^c(\mathbf{x},t) \sum_{\bar{c}} \sum_{i} \omega_i g^{c\bar{c}} \Psi^{\bar{c}}(\mathbf{x}+\mathbf{e}_i,t) \mathbf{e}_i\,\,\,
\end{equation}
to achieve separation of the components~\cite{Shan1993}. This force is defined
as a nearest neighbor interaction between fluid components $c$ and
$\bar{c}$~\cite{Shan1993} and scaled through the choice of the parameter
$g^{c\bar{c}}$.  Here $\Psi^c(\mathbf{x},t)$ is an effective mass, defined as
\begin{equation}
    \label{eq:psifunc}
    \Psi^c(\mathbf{x},t) \equiv \Psi(\rho^c(\mathbf{x},t) ) = 1 - e^{-\rho^c(\mathbf{x},t)/\rho_0}
    \mbox{.}
\end{equation}
The force is applied to the fluid by adding a shift of $\Delta
\mathbf{u}^c(\mathbf{x},t) =\frac{\tau^c
\mathbf{F}^c(\mathbf{x},t)}{\rho^c(\mathbf{x},t)}$ to
$\mathbf{u}^c_{eq}(\mathbf{x},t)$ during collision.  This causes the separation
of fluids and the formation of a diffuse interface between them.  The width of
the interface separating the regions is typically about $5 \Delta
x$~\cite{Frijters2012}, with a small dependence on the interaction strength.

\subsection{The evaporation model}
When the interaction parameter $g^{c\bar{c}}$ in the pseudopotential model is
properly chosen~\cite{Martys1996a}, a separation of components takes place.
Each component will separate into a denser {\it majority} phase of density
$\rho_{ma}$ and a lighter {\it minority} phase of density $\rho_{mi}$,
respectively. 

In order to drive the system out of equilibrium, we impose the density of component $c$
at the boundary sites $\mathbf{x}_H$ to be of constant value $
\rho^c(\mathbf{x}_H,t)= \rho^c_H$ by setting the distribution function of
component $c$ to
\begin{equation}
    f_i^c(\mathbf{x}_H,t) = f_i^\mathrm{eq}\left(\rho_H^c,\mathbf{u}^c_H(\mathbf{x}_H,t)\right),
\end{equation}
in which $\mathbf{u}^c_H(\mathbf{x}_H,t)=0$.
Setting the velocity to zero at the system boundary is in agreement with the idea of an undisturbed large volume outside the system, which does not cause perturbations in the system.
Depending on the ratio of the minority density $\rho^c_{mi}$ and $\rho_H^c$ this induces evaporation or condensation.
Furthermore, \revisedtext{for simplicity, we ensure total mass conservation within the system by setting the density of component $\bar{c}$ as}
\begin{equation}
    \rho^{\bar{c}}(\mathbf{x}_H,t) = \rho^c(\mathbf{x}_H,t-1) + \rho^{\bar{c}}(\mathbf{x}_H,t-1) - \rho^c_H
\end{equation}
and again ensure an undisturbed flow field by setting $\mathbf{u}^{\bar{c}}_H(\mathbf{x}_H,t)=0$,
so that the distribution functions of component $\bar{c}$ at the evaporation boundary sites $\mathbf{x}_H$ become
\begin{eqnarray} 
    f_i^{\bar{c}}(\mathbf{x}_H,t) = f_i^\mathrm{eq}\left(\rho_H^{\bar{c}},\mathbf{u}^{\bar{c}}_H (\mathbf{x}_H,t)\right).
\end{eqnarray}
\revisedtext{We note that our model can be easily extended to situations where mass is not conserved.}
We ensure the equivalence to the open system by mimicking a zero density 
gradient within the infinite volume outside the system.  
\revisedtext{This way forces from the pseudopotential interactions only have an impact in the bulk of the simulation volume and not at the boundary.
    This can be achieved either by using evaporation
    boundaries on periodic sites as well or by using a second layer of
    evaporation boundary sites. Thereby we enforce the same density and a zero
gradient at the system boundary.}

In the case where the set density $\rho_H^c$ is not equal to the equilibrium
minority density $\rho_{mi}^c$, a density gradient in the vapor phase of
component $c$ is formed. This gradient causes component $c$ to diffuse towards
the minimum.

\section{Results and Discussions} \label{sec:results}

\subsection{Diffusion} \label{subsec:diffusion}
In binary fluid mixtures, following Fick's first law, we can write the mass flux of component $c$ as
\begin{equation}
    \mathbf{j}^{c} =  -D^{cc} \nabla \rho^c - D^{c{\bar{c}}}\nabla \rho^{\bar{c}},
    \label{eq:fick}
\end{equation}
where $D^{cc}$, $D^{c{\bar{c}}}$ are the diffusion coefficients.
In the Shan-Chen multicomponent method, the mass flux of component $c$ can be written as~\cite{Shan1996}
\begin{equation}
    \mathbf{j}^{c} = \rho^{c}(\mathbf{U}^c-\mathbf{U}),
    \label{eq:flux}
\end{equation}
where $\mathbf{U}^c$ and $\mathbf{U}$ are macroscopic velocities of component $c$ and the binary mixture, respectively.
The macroscopic velocities are defined as an average of the total momentum before and after each collision~\cite{Shan1995a} as
\begin{eqnarray}
    \mathbf{U} &=& \frac{\rho^c\mathbf{u}^c + \rho^{\bar{c}}\mathbf{u}^{\bar{c}} + \frac{1}{2}(\mathbf{F}^c +\mathbf{F}^{\bar{c}})}{\rho^c+\rho^{\bar{c}}}\mbox{,} \\
    \mathbf{U^c} &=&  \frac{1}{2\rho^c} \left[ \rho^c\mathbf{u}^c +\mathbf{F}^c + \frac{\rho^c(\rho^c \mathbf{u}^c+\rho^{\bar{c}}\mathbf{u}^{\bar{c}}) }{\rho^c+\rho^{\bar{c}}} \right]
    \mbox{.} \label{eq:velw}
\end{eqnarray}
By performing a Chapman-Enskog expansion~\cite{Shan1995a}, \eqnref{flux} can be rewritten to be identical to \eqnref{fick}, with the diffusion coefficients given as~\cite{Shan1996}
\begin{eqnarray}
   \!\!\! D^{cc} &= & c_s^2\left(\tau-\frac{1}{2}\right)\frac{\rho^{\bar{c}}}{\rho^c+\rho^{\bar{c}}} -\frac{c_s^2 \rho^{c} \Psi^{\bar{c}}  g^{\bar{c}c} \Psi^{'c} } {\rho^c+\rho^{\bar{c}}}, \nonumber \\
    \!\!\!D^{c{\bar{c}}} &=& -c_s^2\left(\tau-\frac{1}{2}\right)\frac{\rho^c}{\rho^c+\rho^{\bar{c}}} + \frac{c_s^2\rho^{\bar{c}} \Psi^c g^{c\bar{c}} \Psi^{'\bar{c}} } {\rho^c+\rho^{\bar{c}}} ,
    \label{eq:diffu-t-1}
\end{eqnarray}
with $\Psi^{'c}$ being the spatial derivative of $\Psi^{c}$.
We note that the diffusion is dependent on the symmetric interaction strengths $g^{c\bar{c}}$ and $g^{\bar{c}c}$, as well as the densities of the two components $\rho^c$ and $\rho^{\bar{c}}$.

In the limit of small gradients, we can assume that $\nabla \rho^{\bar{c}}(\mathbf{x},t)  = -\nabla \rho^{c}(\mathbf{x},t) $, with which \eqnref{fick} becomes 
\begin{equation}
    \mathbf{j}^{c} = -D^c \nabla \rho^c,
    \label{eq:fick-simple}
\end{equation}
where  
\begin{equation}
\!\!\!\!\!\!\!\!\!\!\!\!\!\!\!\!    D^c=\biggl[c_s^2(\tau-\frac{1}{2}) 
         -  \frac{c_s^2}{\rho^c+\rho^{\bar{c}}} (\rho^{\bar{c}} \Psi^c g^{c\bar{c}} \Psi^{'\bar{c}} + \rho^{c} \Psi^{\bar{c}}  g^{\bar{c}c} \Psi^{'c} )\biggr].
\end{equation}

\begin{figure}
    \includegraphics{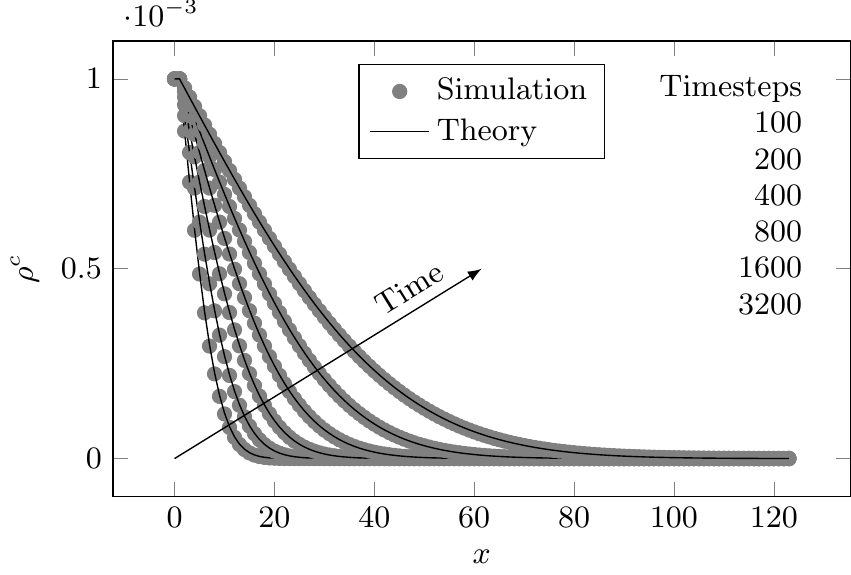}
    \caption{Time evolution of density profiles of component $c$. The system is initially empty of $c$.
    Diffusion allows fluid from an infinite reservoir at $x=0$ to build up gradually changing density profiles in agreement with the analytical solution.
    \label{fig:com}}
\end{figure}

To validate the theoretical analysis above, we investigate the diffusion of a component $c$ into a system filled with another component $\bar{c}$. 
We perform a simulation with a system size of $125\times4\times4$ and fill the system with fluid $\bar{c}$ of density $\rho^{\bar{c}}=0.7$.
A wall of thickness \revisedtext{$h=2$} is placed at $x=125$ and the fluid interaction parameter is set to $g^{c\bar{c}}= g^{\bar{c}c}=3.6$.
We utilize an evaporation boundary at $x=0$ and set the density $\rho^c_H = 0.001$ to ensure a diffusive flow that is undisturbed by convection. 
Then component $c$ diffuses into the system.
Meanwhile, we numerically solve Fick's second law
\begin{eqnarray}
    \frac{\partial \rho^c}{\partial t}=-D^{cc} \Delta \rho^c - D^{c{\bar{c}}}\Delta \rho^{\bar{c}} \nonumber \\
    \frac{\partial \rho^{\bar{c}}}{\partial t}=-D^{\bar{c}\bar{c}} \Delta \rho^{\bar{c}} - D^{{\bar{c}}c}\Delta \rho^{c}
    \label{eq:fick2}
\end{eqnarray}
to describe the space and time dependent density profile of component $c$.
In \figref{com} we compare the lattice Boltzmann simulation results (symbols) with the numerical solution of~\eqnref{fick2} (solid lines).
From the evaporation boundary with density $\rho_H^c$ fluid diffuses into the system.
Being a diffusion process, the rate at which the fluid invades the system is dependent on the density gradient.
The gradient subsequently decreases and fluid distributes itself further into the system, aiming to remove the gradient.
There is a good agreement between simulation results and the numerical solution, as shown in~\figref{com}.

We note that the diffusion equation does not hold at fluid-fluid interfaces~\cite{Shan1996}.  
However, the movement of the interface during evaporation is governed by diffusion of the fluids surrounding it, which we demonstrate as follows.

\subsection{Evaporation of a planar film}\label{sec:film}
\label{subsec:film}

\begin{figure}
    \centering
    \includegraphics{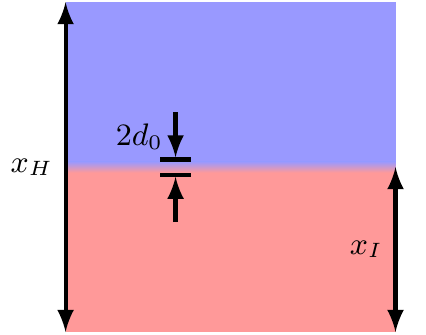}
    \caption{\label{fig:film}(Color online) Schematic representation of the planar film (front view).
             We fill the lower half of the system with fluid $c$ and the upper half with fluid $\overline{c}$ such that a fluid-fluid
             interface forms at $x_I$. The interface thickness is $2d_0$. To drive the evaporation we impose the boundary condition $\rho^{c}(x=x_H) = \rho^{c}_{H}$ 
             at the top of the system. A solid surface with thickness $h=2$ is located at the bottom.
}
\end{figure}

\begin{figure}
    \includegraphics{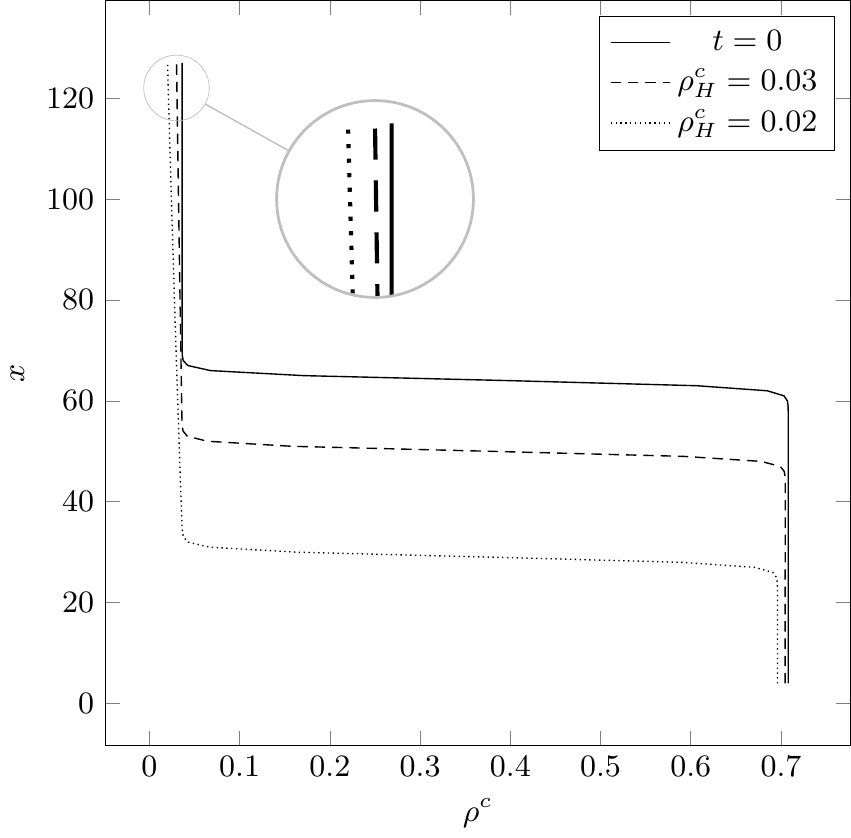}
    \caption{\label{fig:film-den}Density profile of fluid $c$ along the $x$ direction after equilibration (defined as $t=0$, solid line)
             and density profiles at $t=10^6$ timesteps later with boundary densities $\rho^{c}_{H} = 0.03$ (dashed line) and $\rho^{c}_{H} = 0.02$ (dotted line). 
             The magnification depicts of the subtle difference of $\rho^{c}_{H}$ causing a different density gradient and a different time behavior of the moving interface.}
\end{figure}
 
We investigate the evaporation of a planar film sitting on a solid substrate,
as illustrated in ~\figref{film}.  To do so we perform simulations with a
system size of $128 \times 4 \times 4$.  We fill one half of the system with
fluid $c$ and the other half with fluid $\bar{c}$ of equal density 
($\rho^c_{ma}=\rho^{\bar{c}}_{ma} = 0.70$, $\rho^c_{mi}=\rho^{\bar{c}}_{mi} = 0.04$) 
such that a fluid-fluid interface forms at $x_0=64$. We define the position of
the interface $x_I$ as the position of $\rho^c-\rho^{\bar{c}}=0$.  The
interaction strength in~\eqnref{sc} is chosen to be $g^{c\bar{c}} =g^{\bar{c}c}
= 3.6$.  We place a wall of thickness $2$ with simple bounce back boundary
conditions at the bottom, parallel to the interface, while the boundaries
normal to the substrate are periodic.

After equilibration the density of fluid $c$ is constant in both the denser
phase ($\rho^{c}_{ma} \approx 0.704$) and the lighter phase ($\rho^{c}_{mi}
\approx 0.036$), whereas between them a diffuse interface of about $2d_0=5$
lattice units is formed, as shown in the density profile along $x$ direction
in~\figref{film-den} (solid line). We then apply the evaporation boundary condition by setting the density at the 
top boundary $\rho^{c}(x=128)$ to $\rho^c_{H}$.
In \figref{film-den} we show the density profiles along the $x$ direction just after equilibration (solid line) and for
evaporation boundary densities $\rho^{c}_{H} = 0.03$ (dashed line) and
$\rho^{c}_{H} = 0.02$ (dotted line) after $10^6$ subsequent simulation timesteps.  A
density gradient of fluid $c$ is formed in the lighter phase, resulting in
diffusion of fluid $c$ towards the evaporation boundary.  Thus, the interface
position decreases with time.  It decreases faster for a lower evaporation
boundary density $\rho^{c}_{H}$, which indicates that the mass flux increases
with decreasing the evaporation boundary density.

If we assume that the fluid densities in the minority phases vary linearly, \eqnref{fick-simple} becomes
\begin{equation}
  \mathbf{j}^{c} = -\left( D^c (\rho_{mi}-\rho^c_{H}) / (x_H-x_I-d_0) \right) \mathbf{n},
  \label{eq:fick-linear-0}
\end{equation}
where $\mathbf{n}$ is the normal vector of the interface.
The mass flux is approximately proportional to the density difference between the minority density
and the evaporation boundary density $\rho_{mi}-\rho^c_{H}$, which 
allows to control the evaporation rate by varying $\rho^c_{H}$.
\begin{figure}
    \centering
    \includegraphics{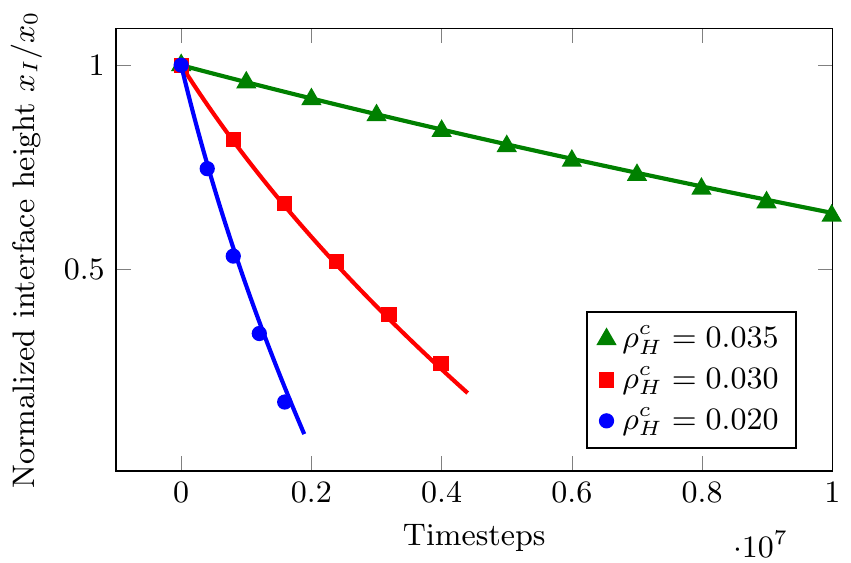}
    \caption{\label{fig:film-ht}(Color online) Interface position as a function of time for different evaporation boundary 
             densities $\rho^{c}_{H} = 0.035$, $\rho^{c}_{H} = 0.03$ and $\rho^{c}_{H} = 0.02$.
             The theoretical prediction~\eqnref{xt} (solid lines) agrees well with the simulation data (symbols)}
\end{figure}

With the assumption that the density profile across the interface is also linear, 
the total mass of fluid $c$ in the system is
\begin{eqnarray}
    M^c(t)  = A \big[ (x_I - d_0) \rho^{c}_{ma} + (x_H - x_I -d_0)\nonumber \\
    (\rho^{c}_{mi}+\rho^c_H)/2 + d_0(\rho^{c}_{ma} - \rho^{c}_{mi}) \big]     \mbox{,}
    \label{eq:mass-1}
\end{eqnarray}
where $A$ is the area of the cross-section. 
From~\eqnref{mass-1}, we can obtain
\begin{equation}
    dM^c/dt = A(\rho^{c}_{ma} -\rho^{c}_{mi}/2- \rho^{c}_{H}/2)  \frac{d x_I}{dt}
    \mbox{.}
    \label{eq:mass_flux-1}
\end{equation}
%
Based on the principle of mass conservation, the time evolution of mass obeys
\begin{equation}
    M^c(t) = M^c(0)- A\int_0^t \mathbf{j}^{c}\cdot \mathbf{n} dt
    \mbox{,} \label{eq:mass-2}
\end{equation}
where $M^c(0)$ is the initial mass of fluid $c$.
From~\eqnref{mass-2}, we can also get the time derivative of the total mass as
\begin{equation}
    dM^c/dt = A|\mathbf{j}^{c}| 
    \mbox{.}
    \label{eq:mass_flux-2}
\end{equation}
By comparing \eqnref{mass_flux-1} and \eqnref{mass_flux-2}, we obtain 
\begin{equation}
    \frac{d x_I}{dt} = \frac{D (\rho^c_{mi}-\rho^c_{H})}{(x_H-x_I-d_0)(\rho^{c}_{ma}-\rho^{c}_{mi}/2-\rho^{c}_{H}/2)}.
    \label{eq:xit}\,\,\,
\end{equation}
We solve~\eqnref{xit} with the initial condition $x_I(t=0)=x_0$ and finally obtain the interface position 
as a function of time
\begin{eqnarray}
   x_I(t) = x_H+d_0 - \big[ (x_H+d_0-x_0)^2 \nonumber \\
   + 2 \frac{D (\rho^c_{mi}-\rho^c_{H}) } {\rho^{c}_{ma} -\rho^{c}_{mi}/2-\rho^{c}_{H}/2} t \big]^{1/2}. 
   \label{eq:xt}
\end{eqnarray}

The simulation results of the time evolution of the interface position for different evaporation boundary densities
$\rho^{c}_{H} = 0.035$, $\rho^{c}_{H} = 0.03$ and $\rho^{c}_{H} = 0.02$ along with our theoretical analysis \eqnref{xt} are presented in \figref{film-ht}. 
We find excellent quantitative agreement between theory and simulation.

\subsection{Evaporation of a freely suspended droplet} \label{subsec:ep}
\begin{figure}
    \vskip3ex
    \includegraphics{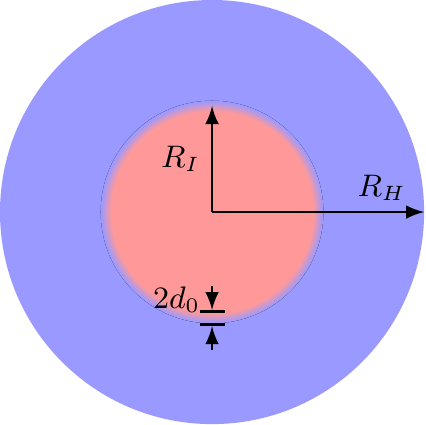}
    \caption{(Color online) Schematic cross-sectional representation of a droplet of radius $R_I$ surrounded by another fluid up to the spherical system boundary $R_H$, where the evaporation boundary condition is imposed.}
    \label{fig:droplet}
\end{figure}
In this section we investigate the evaporation of a freely floating droplet. 
A droplet of component $c$ with a radius of $R_I$ is the center of a spherical system of size $R_H$ and surrounded by component $\bar{c}$, as shown in~\figref{droplet}. 
A spherical evaporation boundary is applied at $R_H$. 
Under the assumption of quasi-static dynamics, \revisedtext{the density profile of component $c$ in the lighter phase satisfies the Laplace equation,
 \begin{equation}
\Delta \mathbf{\rho}^{c}(r) = 0	
\mbox{,}
\end{equation}
where the boundary conditions are
\begin{equation}
 \mathbf{\rho}^{c}(r) |_{r=R_I+d_0} = \rho^c_{mi} 
\end{equation}
and 
\begin{equation}
 \mathbf{\rho}^{c}(r) |_{r=R_H}= \rho_{H}^{c}
 \mbox{.}
\end{equation}
}
\revisedtext{In spherical coordinates, we obtain the analytical solution as } 
\begin{equation}
\!\!   \mathbf{\rho}^{c}(r) = \rho_{H}^{c}- (\rho_{H}^{c}-\rho^c_{mi}) \frac{R_H-r}{R_H-R_I-d_0} \frac{R_I+d_0}{r},\,\,
    \label{eq:fick-linear}
\end{equation}
where $r$ is the distance from the center of a spherical coordinate system, originating at the droplet center.
Inserting~\eqnref{fick-linear} into \eqnref{fick-simple}, we obtain the mass flux as
\begin{equation}
    \mathbf{j}^{c}(r) = - D^c (\rho_{mi}-\rho^c_{H})  \frac{R_H(R_I+d_0)}{(R_H-R_I-d_0)r^2} \mathbf{n}_r,
    \label{eq:flux-drop}
\end{equation}
where $\mathbf{n}_r$ is the normal vector to the droplet interface.
Assuming the density profile across the interface also satisfies Laplace's equation, we obtain the total mass of fluid $c$ in the system as
\begin{widetext}
\begin{eqnarray}
    M^c  & =& \int_{0}^{R_I-d_0} 4\pi r^2 \rho^{c}_{ma} dr + \int_{R_I-d_0}^{R_I+d_0} 4\pi r^2 \left( \rho^c_{mi}- (\rho^c_{mi}-\rho^c_{ma}) \frac{R_I+d_0-r}{2d_0} \frac{R_I-d_0}{r}\right) dr \nonumber
    \\ && +\int_{R_I+d_0}^{R_H} 4\pi r^2 \left( \rho_{H}- (\rho_{H}-\rho^c_{mi}) \frac{R_H-r}{R_H-R_I-d_0} \frac{R_I+d_0}{r}\right) dr     
    \mbox{.}
    \label{eq:drop-mass-1}
\end{eqnarray}
\end{widetext}
We simplify~\eqnref{drop-mass-1} and derive the time derivative of the total mass as\\
\vskip-3ex\begin{multline}
    d M^c / dt = \frac{2\pi}{3}\biggl(
        R_H^2\left(-\rho_H+\rho^c_{mi}\right) \\
        + R_H \left( -2R_I \rho^c_H+2R_I \rho^c_{mi} - 2d_0\rho^c_H+2d_0\rho^c_{mi}\right)\\
        + 6R_I^2 \rho^c_{ma} - 6R_I^2\rho^c_{mi} -2d_0^2\rho^c_{ma} +2d_0^2\rho^c_{mi} 
    \biggr) d R_I/dt.
    \label{eq:mass-df2}
\end{multline}
Based on the principle of mass conservation we have
\begin{equation}
    M^c(t) = M^c(0)- \int_0^t 4\pi r^2\mathbf{j}^c\cdot\mathbf{n}_r dt
    \mbox{,} \label{eq:mass-3}
\end{equation}
where $M^c(0)$ is the total initial mass of fluid $c$ in the system.
From~\eqnref{mass-3} with using~\eqnref{flux-drop}, we also get the time derivative of the total mass as  
\begin{equation}
    dM^c/dt = 4 \pi D^c (\rho_{mi}-\rho^c_{H}) \frac{(R_I+d_0)R_H}{(R_H-R_I-d_0)}
    \mbox{.} \label{eq:mass-df1}
\end{equation}

By comparing \eqnref{mass-df2} and \eqnref{mass-df1} we obtain the time evolution of the droplet radius as
\begin{multline}
    dR_I/dt = 4 \pi D^c (\rho_{mi}-\rho^c_{H}) (R_I+d_0)R_H \\
    \bigg[ (R_H-R_I-d_0)    \frac{2 \pi}{3} \biggl( R_H^{2} \left(-  \rho_{H} +  \rho^c_{mi} \right) \\
    + R_H \left(- 2R_I \rho^c_{H} + 2R_I \rho^c_{mi} - 2 d_0 \rho^c_{H} + 2 d_0 \rho^c_{mi} \right)\\
        + 6 R_I^{2} \rho^c_{ma} - 6 R_I^{2} \rho^c_{mi} - 2 d_0^{2} \rho^c_{ma} + 2 d_0^{2} \rho^c_{mi}\biggr)      \bigg]^{-1}    
        \label{eq:mass-dfsum-sim}
        \mbox{.}
\end{multline}
We solve~\eqnref{mass-dfsum-sim} numerically with a $4th$-order Runge-Kutta algorithm.
For the simulations we initialize a droplet with a radius of $R_0=65$ and densities $\rho^c_{ma} = \rho^{\bar{c}}_{ma}=0.70$, $\rho^c_{mi} = \rho^{\bar{c}}_{mi}=0.04$, in a computational domain of $256^3$.
The densities of fluid $c$ equilibrate to $\rho^c_{ma}\approx 0.712$ inside the droplet and $\rho^c_{mi}\approx0.037$ outside.
We initiate the evaporation by setting the density at the spherical evaporation
boundary $r=R_H$ to $\rho^{c}_{H}=0.03$, $\rho^{c}_{H}=0.025$, and
$\rho^{c}_{H}=0.02$. In~\figref{EPSurfacetension} we compare the analytical
solution~\eqnref{mass-dfsum-sim} (dashed lines) with the simulation results
(symbols). \revisedtext{We note that Shan-Chen models are exposed to spurious vaporisation effects once the droplets become small, i.e. when the diameter is around $5-10$ lattice units. 
\begin{figure}
    \includegraphics{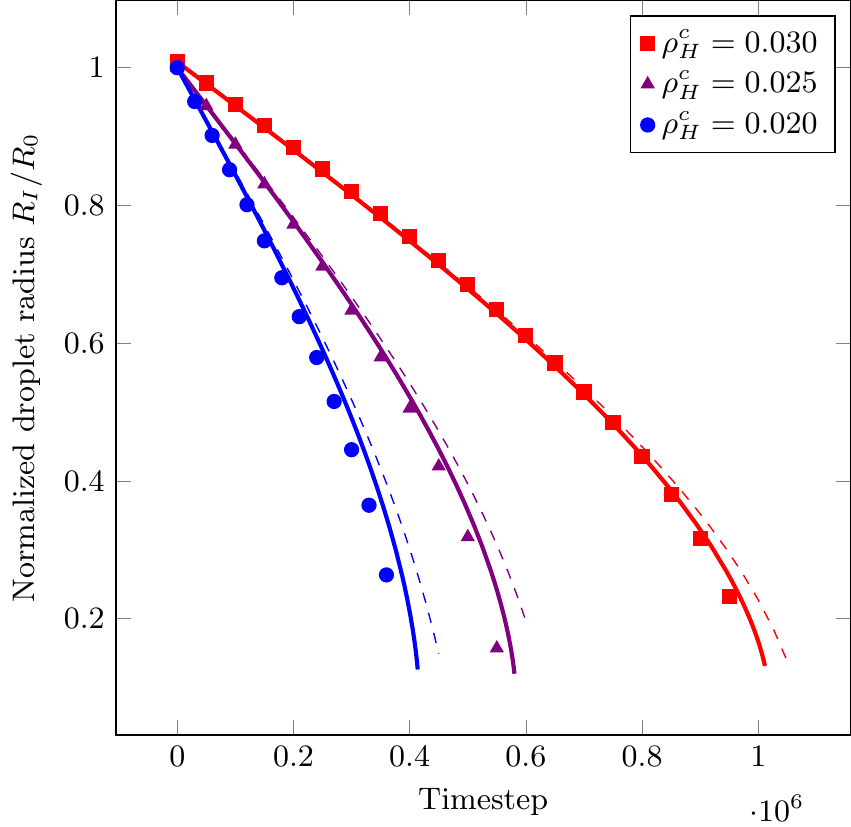}
    \caption{\label{fig:EPSurfacetension}(Color online) Time evolution of the droplet radius for evaporation boundary densities $\rho^{c}_{H} = 0.03$, $\rho^{c}_{H} = 0.025$ and $\rho^{c}_{H} = 0.02$.
             Our theoretical analysis without surface tension~\eqnref{mass-dfsum-sim} (dashed lines) agrees quantitatively with the simulation data (symbols) for large droplet radii and deviates for small droplet radii.
             Our theoretical analysis which includes the surface tension~\eqnref{drop-r-df4} (solid lines) agrees quantitatively well with the simulation data for both large and small droplet radii.
             }
\end{figure}
To avoid the effect of the spurious vaporisation on the analysis, we only use the simulation data when the diameter of the droplet is larger than $20$}
The analytical solution captures the qualitative features of the time evolution of the droplet radius well, 
and quantitatively agrees with the simulation data for the droplet at a larger radius. However, it deviates for small droplet radii. 
This can be explained by the fact that we neglected the effect of surface tension on the droplet evaporation. 
The surface tension induces a Laplace pressure, which is larger when the droplet radius becomes small~\cite{EpsPle50}.
We can take into account this effect as follows:
  
\newcommand{\outside}{\ensuremath{r>R_I}}
\newcommand{\inside}{\ensuremath{r<R_I}}
For a spherical droplet the Young-Laplace equation can be written as
\begin{equation}
    P(\outside,t) = P(\inside,t) - \frac{2\gamma}{R_I},
\end{equation}
where $\gamma$ is the surface tension, $P(\outside,t)$ and $P(\inside,t)$ are the pressures outside and inside the droplet at time $t$, respectively.
We can write the pressure inside the droplet as~\cite{Shan1993}
\begin{equation}
\hspace*{-5ex}P(\inside,t) = c_{s}^{2} (\rho^c_{ma} + \rho^{\bar{c}}_{mi})  +  \frac{c_{s}^{2}}{2} g_{c\bar{c}}\Psi(\rho^c_{ma})\Psi(\rho^{\bar{c}}_{mi}).\,
\end{equation}
For simplification, in the case of $\rho^{\bar{c}}_{mi}\ll \rho^c_{ma}$ we can write the pressure in terms of the leading term as
\begin{equation}
    P(\inside,t)= c_{s}^{2} \rho^c_{ma}.
    \label{eq:pressure-rho}
\end{equation}
The pressure outside the droplet can be treated as constant during evaporation, so that we get 
\begin{eqnarray}
    P(\outside,t) &=& P(\inside,t=0) - \frac{2\gamma}{R_0} \nonumber \\
    &=& P(\inside,t) -\frac{2\gamma}{R_I}.
    \label{eq:pre-pre}
\end{eqnarray}
By inserting \eqnref{pressure-rho} into \eqnref{pre-pre}, we obtain the majority density of fluid $c$ inside the droplet as
\begin{equation}
    \rho^c_{ma}(t) = \rho^{c}_{ma}(t=0) - \frac{2\gamma}{c_{s}^{2}} \left(\frac{1}{R_0} - \frac{1}{R_I}\right).
    \label{eq:rho_ma}
\end{equation}

The minority density of fluid $c$ outside the droplet can be treated as proportional to the majority density of fluid $c$ inside the droplet~\cite{EpsPle50}. Thus we obtain

\begin{equation}
\!\!\!\!\!    \rho^{c}_{mi}(t) = \rho^{c}_{mi}(t=0) \frac{\rho^c_{ma}(t=0) - \frac{2\gamma}{c_{s}^{2}} (\frac{1}{R_0} - \frac{1}{R_I}) }{\rho^{c}_{ma}(t=0)}.\,\,\,
    \label{eq:rho_mi}
\end{equation}
For brevity, we denote $\rho^{c}_{ma}(t=0)$ as $\rho^{c}_{ma,0}$ and $\rho^{c}_{mi}(t=0) $ as $\rho^{c}_{mi,0}$.
We insert~\eqnref{rho_ma} and~\eqnref{rho_mi} into~\eqnref{drop-mass-1}, and after some manipulations, we finally obtain the time derivative of the droplet mass as
\begin{widetext}
    \begin{eqnarray}
        dM^c/dt  & =& \frac{2 \pi}{3} \biggl( (-R_H^{2}-2R_IR_H-2 R_H d_0)\rho^c_{H}\nonumber \\
                 &+& (R_H^{2}  + 2R_IR_H   + 2 R_Hd_0 - 6 R_I^{2}   + 2d^{2})\left( \rho^{c}_{mi,0} \frac{\rho^c_{ma,0} - \frac{2\gamma}{c_{s}^{2}} (\frac{1}{R_0} - \frac{1}{R_I}) }{\rho^{c}_{ma,0}}\right)\nonumber \\
                 &+& (R_IR_H^{2}  + d R_H^{2}+ R_I^{2}R_H   + 2 R_I R_Hd_0   + d_0^{2}R_H - 2 R_I^{3}   + 2 R_I d^{2})\left( \frac{-2\gamma \rho^{c}_{mi,0}}{c_{s}^{2}\rho^{c}_{ma,0}} \frac{1}{R_{I}^{2}}\right)\nonumber \\
     &+& (6 R_I^{2} - 2 d^{2}) (\rho^{c}_{ma,0} - \frac{2\gamma}{c_{s}^{2}} (\frac{1}{R_0} - \frac{1}{R_I}) )+ (2 R_I^{3} - 2 R_I d^{2}) (\frac{-2\gamma}{c_{s}^{2}R_{I}^{2}})\biggr)dR_I/dt.
        \label{eq:drop-mass-df4}
    \end{eqnarray}
\end{widetext}

We compare \eqnref{drop-mass-df4} with \eqnref{mass-df1} and get the equation for $dR_I/dt$ including the effect of surface tension as
\begin{widetext}
    \begin{eqnarray}
    dR_I/dt  & =& 4 \pi D^c (\rho_{mi}-\rho^c_{H}) (R_I+d_0)R_H \bigg[ (R_H-R_I-d_0)  \frac{2 \pi}{3} \biggl( (-R_H^{2}-2R_IR_H-2 R_H d_0)\rho^c_{H} \nonumber \\
        &+& (R_H^{2}  + 2R_IR_H   + 2 R_Hd_0 - 6 R_I^{2}   + 2d^{2})\left( \rho^{c}_{mi,0} \frac{\rho^c_{ma,0} - \frac{2\gamma}{c_{s}^{2}} (\frac{1}{R_0} - \frac{1}{R_I}) }{\rho^{c}_{ma,0}}\right)\nonumber \\
             &+& (R_IR_H^{2}  + d R_H^{2}+ R_I^{2}R_H   + 2 R_I R_Hd_0   + d_0^{2}R_H - 2 R_I^{3}   + 2 R_I d^{2})\left( \frac{-2\gamma \rho^{c}_{mi,0}}{c_{s}^{2}\rho^{c}_{ma,0}} \frac{1}{R_{I}^{2}}\right)\nonumber \\
        &+& (6 R_I^{2} - 2 d^{2}) (\rho^{c}_{ma,0} - \frac{2\gamma}{c_{s}^{2}} (\frac{1}{R_0} - \frac{1}{R_I}) )+ (2 R_I^{3} - 2 R_I d^{2}) (\frac{-2\gamma}{c_{s}^{2}R_{I}^{2}})\biggr) \bigg]^{-1}. 
        \label{eq:drop-r-df4}
    \end{eqnarray}
\end{widetext}
We solve \eqnref{drop-r-df4} numerically and compare the theoretical prediction
with simulation data in~\figref{EPSurfacetension}.  The theoretical analysis
including the effect of surface tension (solid lines) agrees quantitatively
well with the simulation data (symbols) for both large and small droplet radii.
Thus, we confirm that the effect of surface tension becomes significant when
the droplets become small and must not be neglected. This result is of
particular importance for lattice Boltzmann simulations of evaporating droplets
since the typical number of lattice nodes available to resolve the radius of a
single droplet is often limited. This holds in particular for systems involving
a large number of droplets.

\section{Conclusion}
\label{sec:final}
We presented a diffusion dominated evaporation model using the popular
pseudopotential multicomponent lattice Boltzmann method introduced by Shan and
Chen. The evaporation is induced by imposing the density of one component at the system boundary while
ensuring total mass conservation, which causes diffusion of components driven by a density gradient.
The diffusion coefficients depend on the densities of the fluids as well as the interaction strength parameters of the Shan-Chen model. With the analytically determined diffusion coefficients we confirm that the diffusion obeys Fick's law.

We derived a theoretical model for the time evolution of the interface position of an evaporating planar film under the quasi-static assumption.
Our theoretical model predicts that the evaporation flux is proportional to the density difference between the minority density
and the evaporation boundary density $\rho_{mi}-\rho^c_{H}$, while the time evolution of the interface position obeys the expected $t^{0.5}$ law. 
We then carried out simulations which are in good quantitative agreement with our analytical model.

\revisedtext{
Furthermore, we derived analytical models describing the evaporation of a floating droplet surrounded by another fluid,
as an extension of the famous Epstein-Plesset theory~\cite{EpsPle50}. While the original publication assumes an infinite system, we extended the model towards a finite system size.
}
We demonstrate a good agreement between theory and simulation if one takes into account the effect of surface tension causing a high Laplace pressure and an increased evaporation rate in the case of small droplet radii.

As an outlook we note that our method is not only suitable to simulate evaporating fluids, but that it is straightforward to apply it to investigate the condensation of droplets. 
Therefore, our method can be a powerful tool for exploring both evaporation and condensation processes in complex fluidic systems. 

\begin{acknowledgments}
    D.~Hessling and Q.~Xie contributed equally to this work.  Financial support
from the Materials innovation institute (M2i, www.m2i.nl,  project number M61.2.12454b), Oc\'e-Technologies B.V.\ and NWO/STW (STW project number 13291) as
well as the allocation of computing time at the High Performance Computing
Center Stuttgart and the J\"ulich Supercomputing Centre are highly
acknowledged.  We thank D.\ Lohse and I.\ Jimidar for fruitful discussions.
\end{acknowledgments}

\end{document}